\documentclass[letterpaper]{article} 
\usepackage[]{aaai23}  
\usepackage{times}  
\usepackage{helvet}  
\usepackage{courier}  
\usepackage[hyphens]{url}  
\usepackage{graphicx} 
\usepackage{natbib}  
\usepackage{caption} 
\usepackage{algorithm}
\usepackage{algorithmic}

\usepackage{amsmath}
\usepackage{amssymb}

\usepackage{multirow}

\usepackage{caption}
\usepackage{subcaption}

\usepackage{pifont}

\usepackage{verbatim}

\usepackage{newfloat}
\usepackage{listings}
\DeclareCaptionStyle{ruled}{labelfont=normalfont,labelsep=colon,strut=off} 
\floatstyle{ruled}
\newfloat{listing}{tb}{lst}{}
\floatname{listing}{Listing}
\nocopyright

\title{Self-supervised Graph Learning for Long-tailed Cognitive Diagnosis}
\author {
    Shanshan Wang\textsuperscript{\rm 1},
    Zhen Zeng \textsuperscript{\rm 1},
    Xun Yang \textsuperscript{\rm 2},
    Xingyi Zhang \textsuperscript{\rm 1}
}
\affiliations {
    \textsuperscript{\rm 1} Anhui University\\
    \textsuperscript{\rm 2} University of Science and Technology of China\\

}

\begin{document}

\maketitle

\begin{abstract}
Cognitive diagnosis is a fundamental yet critical research task in the field of intelligent education, which aims to discover the proficiency level of different students on specific knowledge concepts. Despite the effectiveness of existing efforts, previous methods always considered the mastery level on the whole students, so they still suffer from the Long Tail Effect. A large number of students who have sparse data are performed poorly in the model. To relieve the situation, we proposed a \textbf{S}elf-supervised \textbf{C}ognitive \textbf{D}iagnosis~(SCD) framework which leverages the self-supervised manner to assist the graph-based cognitive diagnosis, then the performance on those students with sparse data can be improved.
Specifically, we came up with a graph confusion method that drops edges under some special rules to generate different sparse views of the graph.
By maximizing the consistency of the representation on the same node under different views, the model could be more focused on long-tailed students.
Additionally, we proposed an importance-based view generation rule to improve the influence of long-tailed students. Extensive experiments on real-world datasets show the effectiveness of our approach, especially on the students with sparse data.



\end{abstract}

\section{Introduction}
Cognitive diagnosis~(CD) is an essential part of the intelligent education system~\cite{liu2021towards,anderson2014engaging}.
Further works in intelligent education, \textit{i.e.}, knowledge tracking~\cite{piech2015deep}, exercise recommendation~\cite{wu2020exercise} are based on the results of cognitive diagnosis.
In the cognitive diagnosis system, there are a series of students, exercises, and knowledge concepts.
For a given student, the cognitive diagnosis task aims to predict the knowledge mastery level on knowledge concepts based on his historical interaction records~\cite{lord1952theory,wang2020neural}.

\begin{figure}[t]
    \centering
    \includegraphics[width=0.5\textwidth]{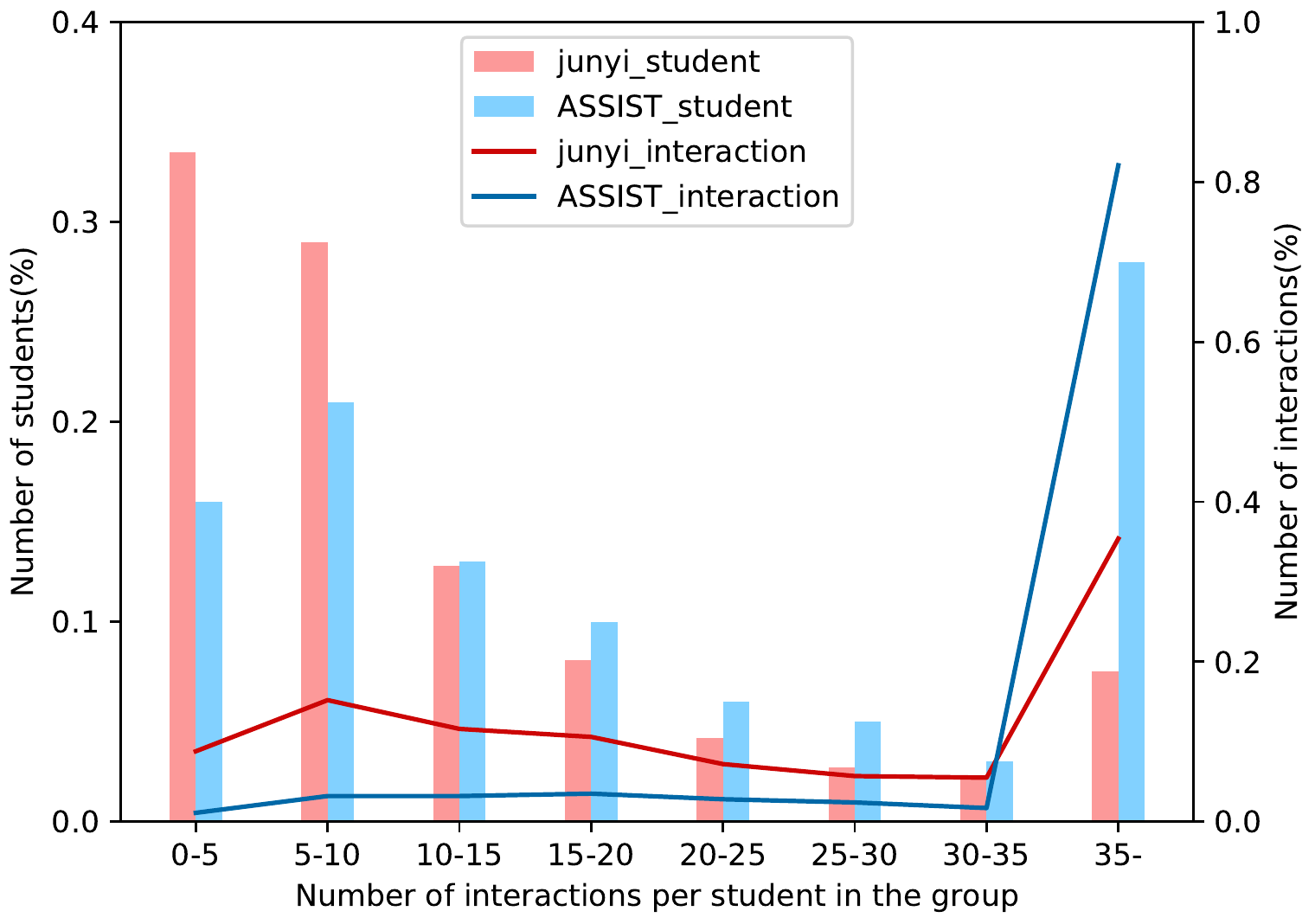}
    \caption{Statistics of the number of students and interactions grouping by interaction records. The horizontal coordinate is the number of interactions per student in the group. The left vertical coordinates indicate the percentage of students in each group to the total students. And the right vertical coordinate indicates the percentage of interactions of all students in each group to the total interactions.}
    \label{students}
    \end{figure}

There has been a great development on cognitive diagnosis,
such as linear model ITR~\cite{lord1952theory}, MIRT~\cite{reckase2009multidimensional},
and neural network-based model NCD~\cite{wang2020neural}.
Recent work RCD~\cite{gao2021rcd} explored the application of graph convolutional networks~(GCN) on cognitive diagnosis and achieved the state-of-the-art results.
However, existing efforts are optimized based on the whole students.
According to our research, there is a serious long-tailed problem in student-exercise interactions.
Statistics of two real-word datasets, \textit{i.e.}, junyi~\cite{chang2015junyi} and ASSIST~\cite{feng2009ASSIST}, are showed in figure~\ref{students}.
Specifically, a minority students have the vast majority interaction records.
Existing models based on overall optimization will favor these students and ignore the long-tailed students.

In order to address the long-tailed problem in cognitive diagnosis, the long-tailed nodes should improve their 
influence and have no impact on other nodes.
Inspired by the SGL~\cite{wu2021self}, we aim to generate sparse views of the original student-exercise interaction graph.
In the specially designed sparse views, the graph dropped edges based on the degree of nodes, so that the long-tailed nodes could achieve a stronger influence.
Additionally, we use the sparse views as an auxiliary self-supervised task, the information discarded in the process
of edge dropout will have no impact on other nodes.
Specifically, we construct a graph contrastive learning for cognitive diagnosis, which provides self-supervised signals by performing self-discrimination of nodes on sparse views. The whole process can be summarized as following.
Firstly, in the graph constructed by student-exercise interaction records, we build the edge importance based on the degree of nodes.
Then we generate different sparse views for the interaction graph according to these edge importance.
Different from the view generation rules in previous self-supervised graph learning work~\cite{wu2021self,zhu2021graph}, long-tailed nodes have greater weight in our generated views.
By maximizing the consistency of the representation on the same node under different views, the model could be more focused on long-tailed students.

Our key contributions are summarized threefold:
\begin{itemize}
\item We propose a new cognitive diagnosis framework SCD based on self-supervised graph learning, which leverages the auxiliary self-supervised signals to alleviate the data sparse problem.
\item In SCD, we design an importance-based view generation rule to enhance the influence of long-tailed students.
\item We have conducted extensive experiments on real-world datasets to validate the effectiveness of our SCD, especially on long-tailed students.
\end{itemize}

\section{Related Work}

\subsection{Cognitive Diagnosis}
Cognitive diagnosis~(CD) is the first step of intelligent education. 
Existing CD methods mainly depend on the assumption that the cognitive states of the student maintain stability in a stable scene.
IRT~\cite{lord1952theory} is considered a milestone of CD, it predicted the accuracy of the student through a manually designed function.
Then, MIRT~\cite{reckase2009multidimensional} extended the latent features of students and exercised in IRT to multi-dimensions.
However, the performance of these traditional models is limited by the design of functions.
To overcome this limitation, NCD~\cite{wang2020neural} leveraged the neural networks to model the interaction between students and exercises.
RCD~\cite{gao2021rcd} first modeled the relationships between knowledge concepts and brought in the graph models to cognitive diagnosis.
Another mainstream foundational paradigm is DINA~\cite{de2009dina}. It directly modeled the relationship between students and knowledge concepts, and took the slip and guessing into account. 
An improved model of DINA is FuzzyCDM~\cite{liu2018fuzzy}, which introduced the fuzzy logic control and realized that the knowledge mastery state could keep a constant value.
Existing CD models have achieved good performance for overall students.
However, none of them takes the long-tailed problem into account.

\subsection{Graph Representation Learning}
Graph-based representation learning has become a popular topic due to the ability of capturing interactions between data. 
GNN~\cite{scarselli2008graph,zhou2020graph} first introduced the information propagation mechanism of neural networks into graph embedding.
GNN-based models had shown great success, such as GCN~\cite{kipf2016gcn} and graphsage~\cite{hamilton2017graphsage}.
GAT~\cite{velivckovic2017gat} introduced the attention mechanism to capture the interactions between neighboring nodes.
Recently, the methods~(\textit{e.g.}, lightGCN~\cite{he2020lightgcn}) of message passing on heterogeneous graphs are explored.
Heterogeneous graphs could represent many real-world interactions, such as the student-exercise relationship in cognitive diagnostics.
However, except RCD~\cite{gao2021rcd}, there is few related work on cognitive diagnosis.

\subsection{Self-supervised Learning}
Self-supervised learning could alleviate the problem of sparse labeling. 
Research on self-supervised learning can be divided into three categories: context-based, temporal-based, and contrastive-based.
Context-based and temporal-based learning are always suitable for Euclidean data, such as text~\cite{mikolov2013efficient,devlin2018bert}, images~\cite{doersch2015unsupervised,noroozi2016unsupervised,pathak2016context}, videos~\cite{sermanet2018time,wang2015unsupervised,misra2016shuffle}, \textit{etc}.
Contrastive-based learning is usually applied on non-Euclidean data such as graph structure data.
Graph contrastive learning~\cite{hjelm2018learning,oord2018representation,zhu2021graph} constructs two different views for the same graph structure, and uses contrastive loss constraints to capture the consistency of feature representations under different views.
Benefiting from its good performance in data sparseness, graph contrastive learning has achieved good performance in recommendation systems~\cite{wu2021self,xia2022hypergraph}.
Inspired by it, to address the long-tailed problem in cognitive diagnosis, we constructed a suitable self-supervised model to learn student and exercise features.

\section{Preliminaries}
We formally define the cognitive diagnosis task.
To apply graph contrastive learning on cognitive diagnostic task, we construct a cognitive diagnostic relationship graph.
\subsection{Problem Statement}
Let $S=\left\{s_{1}, s_{2}, \ldots,
s_{M}\right\}$, $E=\left\{e_{1}, e_{2}, \ldots,
e_{N}\right\}$ and $C=\left\{c_{1}, c_{2}, \ldots, c_{K}\right\}$ be the set of students, exercises and knowledge concepts, respectively.
Q-matrix $Q \in \{0,1\}^{N \times K}$ which contains the relation of exercises with concepts is usually labeled by experts.
$Q_{i,j}=1$ denotes that exercise $e_{i}$ is related to knowledge concept $c_{j}$ and the reverse $Q_{i,j}=0$.
The response records of the students are given in the form of triplet $(s,e,t)$, where $s \in S, e \in E$, and $t$ is the score that the student $s$ got on exercise $e$.
$R$ is the set of response records.
Given the students' response records set $R$ and Q-matrix $Q$, the goal of cognitive diagnosis task is to diagnose students' proficiency on knowledge concepts.

\begin{figure*}[t]
    \centering
    \includegraphics[width=1\textwidth]{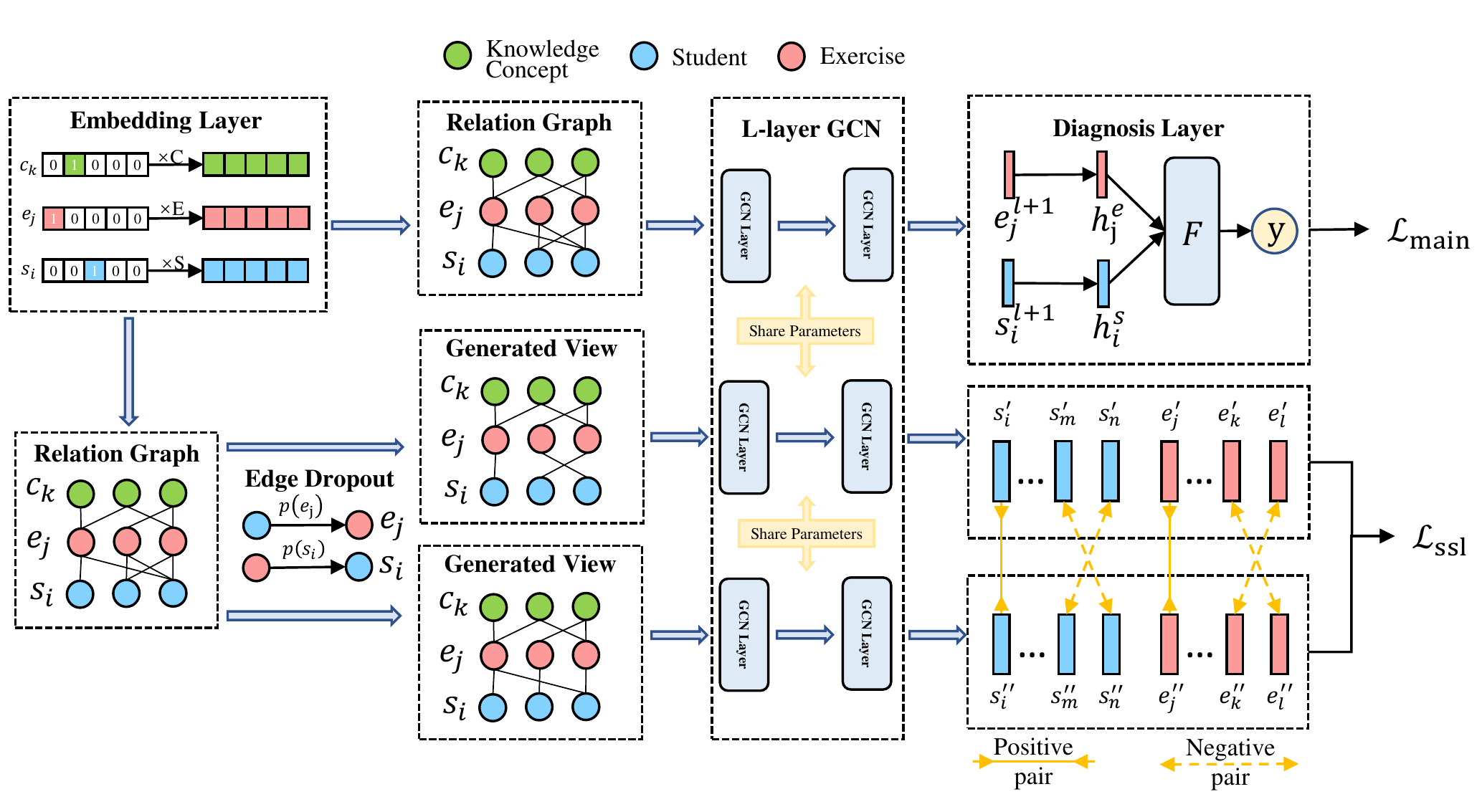}
    \caption{Overview of SCD. The upper part shows the workflow of the cognitive diagnosis main task. And the bottom illustrates the workflow of the self-supervised graph learning task. The main task uses the original relation graph to predict student performance, and the self-supervised graph learning task uses the generated view from edge dropout for contrastive learning. The two tasks share the graph convolutional network.}
    \label{model}
    \end{figure*} 

\subsection{Relation Graph}
Following the student-exercise-concept relation graph proposed in RCD~\cite{gao2021rcd},
we construct a relation graph applicable to graph contrastive learning.
The knowledge concept nodes $C$ act as intermediate nodes to associate exercises which have the same knowledge concept.
Two subgraphs are constructed, which are the student-exercise interaction subgraph $\mathcal{G}_{s e}$ and the exercise-concept relation subgraph $\mathcal{G}_{e c}$.
\subsubsection{Student-Exercise Interaction Subgraph}
We model the interaction record of students and exercises as a bipartite graph $\mathcal{G}_{s e}=(\mathcal{V}_{s e}, \mathcal{R}_{s e})$, where the node set $\mathcal{V}_{s e}=S \cup E$ involves all students and exercise, and edge $r_{s e} \in R_{s e}$ represents student $s$ has answered exercises $e$.

\subsubsection{Exercise-Concept Relation Subgraph}
The construction of the exercises-concept relation graph relies on the Q-matrix.
We denote it as $\mathcal{G}_{e c}=(\mathcal{V}_{e c}, \mathcal{R}_{e c})$,
where the node set $\mathcal{V}_{e c}=E \cup C$ is the union of exercises and knowledge concept,
and edge set $\mathcal{R}_{e c}$ is generated from the Q-matrix.
As the above $r_{e c}\in R_{e c}$ indicates that exercise $e$ involves the knowledge concept $c$.

\section{Methodology}
In this paper, we propose a Self-supervised graph learning for Cognitive Diagnosis~(SCD), which assists main supervised task by self-supervised signals. First, the task of our method is briefly shown. Then we present the main task and self-supervised task separately.
\subsection{Overview}
As shown in Figure~\ref{model}, our SCD model can be divided into two tasks: the main supervised task and the auxiliary self-supervised task.
Two tasks share the GCN model.
The main task is a common graph-based cognitive diagnostic task which leverages the student embeddings and exercise embeddings from the GCN output to make predictions on student scores.
For self-supervised graph learning, we constructed two different views based on the relational graph by the proposed rule of dropping edges.
By maximizing the consistency of the same node on two generated views and minimizing that of different nodes, we could enable GCN to be more focused on long-tailed nodes.

\subsection{Embedding Layer}
Embedding layer encodes the students, exercises, and knowledge concepts into embedding vectors.
Let $S=\mathbb{R}^{M \times d}$, $E=\mathbb{R}^{N \times d}$ and $C=\mathbb{R}^{K \times d}$ be trainable matrix, where $M$, $N$ and $K$ denotes the number of students, exercises and knowledge concepts respectively. $d$ is the dimension of the embedding vector, which is usually set to the same size with the number of knowledge concepts.
Student $s_{i}$ is encoded as one-hot vector. 
We can get its embedding vector $s_{i}^{1}$ by multiplying one-hot vector $s_{i}$ with trainable matrix $S$.
Similarly, the embedding vector of the exercise $e_{j}$ and knowledge concept $c_{k}$ can be obtained in the same way.
The embedding Later can be modeled as:
\begin{equation}
	s_{i}^{1}=s_{i}^{T} S,\
    e_{j}^{1}=e_{j}^{T} E,\
    c_{k}^{1}=c_{k}^{T} C,
\end{equation}
where $s_{i}^{1}, e_{j}^{1}, c_{k}^{1} \in \mathbb{R}^{d}$, $s_{i} \in \{0, 1\}^{M}$, $e_{j} \in \{0, 1\}^{N}$ and $c_{k} \in \{0, 1\}^{K}$. 

\subsection{GCN Model}
The input of the first layer in GCN model is the output of the embedding layer.
GCN is a process of neighborhood cluster based on graph structure.
The representation of each node is the cluster of the neighbor representations in the previous layer.
As shown in figure~\ref{model}, the outputs of the GCN based on the original relational graph and the generated views are leveraged for the cognitive diagnosis main task and the self-supervised learning task respectively.
Let $s_{i}^{l}, e_{j}^{l}$ and $c_{k}^{l}$ be the input student, exercise and knowledge concept embedding victors of the $l$-th GCN layer respectively.
$s_{i}^{l+1}, e_{j}^{l+1}$ and $c_{k}^{l+1}$ are updated after the neighbor aggregation.
As for student $e_{i}$, the embedding is updated in layer $l$ by aggregating the embeddings of the exercise nodes neighbored:
\begin{align}
	s_{i}^{l+1}=\sum_{j \in N_{s_{i}}^{e}} \alpha_{i j}^{l} e_{j}^{l} + s_{i}^{l},
\end{align}
where $N_{s_{i}}^{e}$ are exercises that student $s_{i}$ interacted, $s_{i}^{l}$ is residual connection and $\alpha_{i j}^{l}$ is the node-level attention weight.
The attention weight can be calculated as:
\begin{align}
\label{calalpha}
	\alpha_{i j}^{l}=\operatorname{softmax}(F_{s e}[s_{i}^{l},e_{j}^{l}]),
\end{align}
where $F_{s e}$ represents a full connection layer and $[\cdot]$ is concatenation operation.

The embeddings of students and knowledge concepts are updated in the same way.
Stacking the $l$-layer GCN, the embedding vectors $s^{l+1},e^{l+1},c^{l+1}$ can be obtained incorporating the relation information. 

\subsection{Diagnosis Layer}
Let $h_{i}^{s} \in (0,1)^{K}$ be the mastery level of student $s_{i}$ on each knowledge concept, and $h_{i}^{e} \in (0,1)^{K}$ be the difficulty of exercise $s_{j}$.
After obtaining the embedding vectors of the exercises and students, we can diagnose the cognitive state of students and the difficulty of the exercises through the diagnosis layer:
\begin{align}
	h_{i}^{s}&=\sigma(F_{s}(s_{i}^{l+1})),\\
	h_{j}^{e}&=\sigma(F_{e}(e_{j}^{l+1})),
\end{align}

To verify the correctness of the diagnosis, a prediction function is used to predict the scores of students:
\begin{align}
	y_{i j}=\frac{1}{|N_{e_{j}}^c|}\sum_{k \in N_{e_{j}}^{c}}c_{k}\cdot\sigma(F_{predict}(h_{i}^{s}-h_{j}^{e})),
\end{align}
where $y_{i j}$ is the accuracy that the student $s_{i}$ answered the exercise $e_{j}$.
$N_{e_{j}}^c$ is the set of knowledge concepts involved in exercise $e_{j}$.
$|\cdot|$ is the counting symbol,
and $c_{k}$ is the one-hot code of knowledge concepts.
The loss function of main task is conventional cross entropy loss between predicted score $y$ and true scores $r$.
\begin{align}
    \mathcal{L}_{main}=-\sum_{i}(r_{i} \log y_{i}+(1-r_{i}) \log (1-y_{i})).
\end{align}

\subsection{Self-Supervised Graph Learning}

As shown in figure~\ref{model}, self-supervised graph learning is added to the main task as an auxiliary branch.
It helps GCN to better capture the interaction of sparse nodes and provides self-supervised signals for the model training.
Two different views are generated for the relational graph by dropping edges, which have fewer edges but retain as much important graph structure information as possible.
The detailed process of generating views is described below:

\begin{figure}[t]
    \centering
    \includegraphics[width=0.5\textwidth]{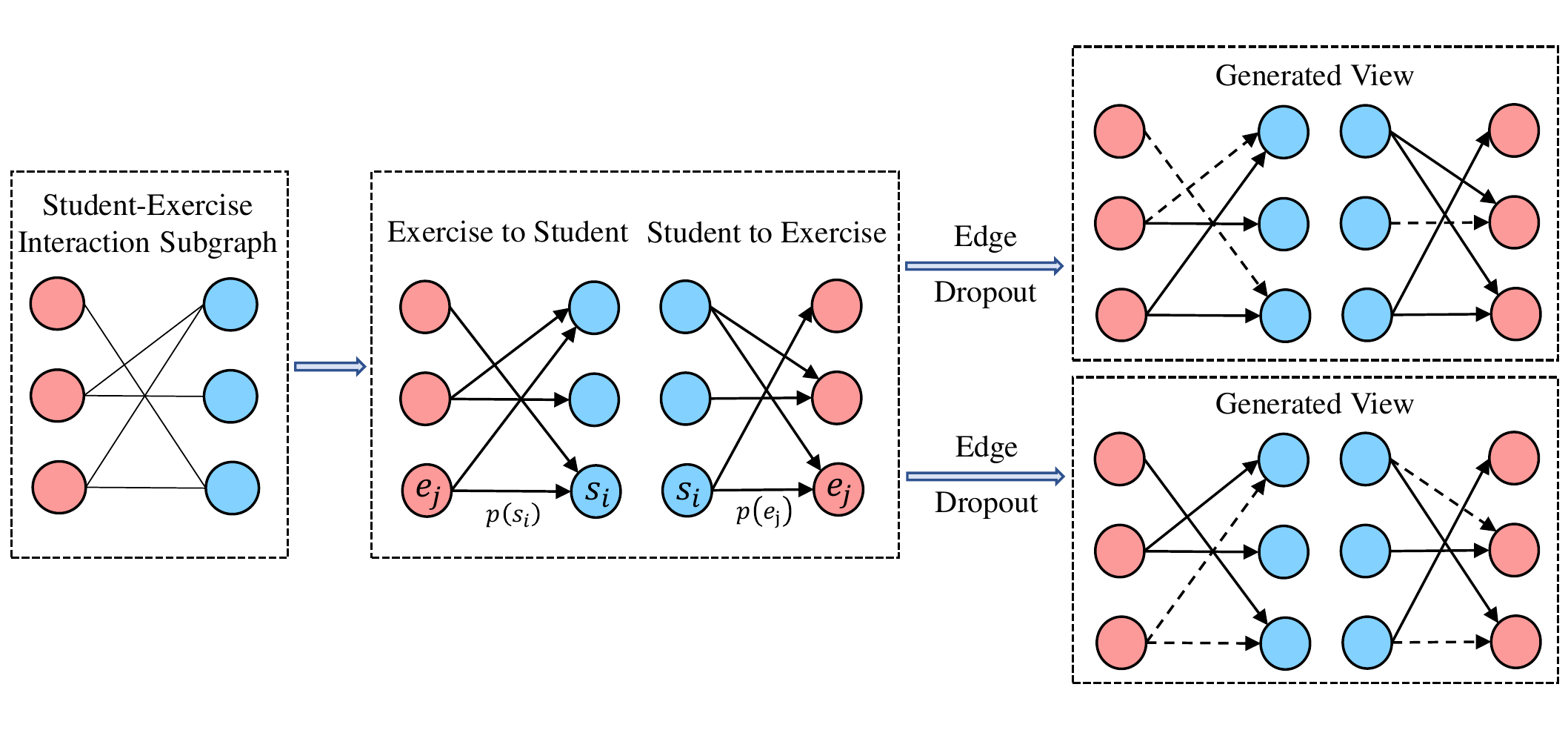}
    \caption{The detailed process of edge dropout. The student-exercise interaction subgraph is split into two directed graphs. The retained probability of the exercise-to-student edge is generated based on the degree of the student node, and does the same for the student-to-exercise edge.}
    \label{model3}
\end{figure} 

\subsubsection{Edge Dropout}
To make GCN focus on long-tailed nodes and prevent them from being biased toward nodes with high-degree, we generate sparse views for the original graph.
In particular, during the generation process, we remove some of the edges connected to high-degree nodes, while keeping the edges of the long-tailed nodes to prevent them from becoming isolated nodes.
Considering the edge on which the degrees of two nodes are quite different, 
retaining such edges would result in generating views that are not sparse enough, while deleting them would produce isolated nodes.
To address this problem, we divide each edge into two directed edges based on the message passing procedure of GCN and process them separately.

Figure~\ref{model3} shows the process of edge dropout.
The target of edge dropout is student-exercise interaction edges, so here only present the student-exercise interaction subgraph.
First we split the student-exercise interaction subgraph into two directed bipartite graphs: exercise to student and student to exercise.
The former one contains only exercises to student edge and is used in the GCN layer for student aggregation.
The latter one has opposite edges and is used for exercise aggregation.
Let $d(s_{i})$ be indegree of $s_{i}$.
For each edge whose vertex is $s_{i}$ in exercise to student subgraph, we calculate their importance $t(s_{i})$ by:
\begin{align}
    \label{tsi}
    t(s_{i})=\frac{k}{ln(d(s_{i})+\theta)},
\end{align}
where $k$ is a hyperparameter which controls the overall probability of edge dropout,
$\theta$ is a small positive value to avoid the numerator being $0$.
The probability $p(s_{i})$ of retaining each edge is calculated by $t(s_{i})$:
\begin{equation}
    \label{psi}
    p(s_{i})=
    \begin{cases}
    p_{min},  & \text{if $t(s_{i}) \le p_{min}$} \\
    t(s_{i}), & \text{if $p_{min} < t(s_{i}) \le 1$} \\
    1, & \text{if $1 < t(s_{i})$}
    \end{cases}
\end{equation}
where $p_{min}$ is the hyperparameter to control the minimum probability that each edge is retained to avoid the generated view from discarding too much information.
In exercise to student subgraph, we generate the retention probabilities for each edge by the same approach as in Eq.~(\ref{tsi}, \ref{psi}):
\begin{gather}
    \label{tpei}
    t(e_{i})=\frac{k}{ln(d(e_{i})+\theta)}, \\
    p(e_{i})=
    \begin{cases}
    p_{min},  & \text{if $t(e_{i}) \le p_{min}$} \\
    t(e_{i}), & \text{if $p_{min} < t(e_{i}) \le 1$} \\
    1, & \text{if $1 < t(e_{i})$}
    \end{cases}
\end{gather}

Based on the probability $p$ for each edge to be retained, the different views of the student-exercise interaction subgraph can be generated.
\subsubsection{Contrastive learning}
In each epoch, two views are generated for the student-exercise interaction subgraph and replace the original subgraph in the relationship graph.
By performing neighbor aggregation on the generated relational graph, the  representations of the same node on different views can be gotten:
$s_{i}^{\prime}, s_{i}^{\prime \prime}$ for $s_{i}$ and $e_{j}^{\prime}, e_{j}^{\prime \prime}$ for $e_{j}$.
We treat the representations of the same node under different views as being positive pairs~(\textit{i.e.}, $(s_{i}^{\prime},s_{i}^{\prime \prime})$),
and the representations of different nodes as negative pairs~(\textit{i.e.}., $(s_{i}^{\prime},s_{j}^{\prime \prime}),i \ne j$).
To maximize the consistency of the positive class samples while minimizing the negative class samples,
we adopt contrastive loss adjusted from InfoNCE~\cite{oord2018representation} as self-supervised loss:
\begin{gather}
    \label{ssl}
    \mathcal{L}_{ssl}^{s}=\frac{1}{M}\sum_{s_{i} \in S}-\log \frac{\exp (sim(\mathbf{s}_{i}^{\prime}, \mathbf{s}_{i}^{\prime \prime}) / \tau)}{\sum_{s_{j} \in S,i \ne j} \exp (sim(\mathbf{s}_{i}^{\prime}, \mathbf{s}_{j}^{\prime \prime}) / \tau)}, \\
    \mathcal{L}_{ssl}^{e}=\frac{1}{N}\sum_{e_{i} \in E}-\log \frac{\exp (sim(\mathbf{e}_{i}^{\prime}, \mathbf{e}_{i}^{\prime \prime}) / \tau)}{\sum_{e_{j} \in E,i \ne j} \exp (sim(\mathbf{e}_{i}^{\prime}, \mathbf{e}_{j}^{\prime \prime}) / \tau)}, \\
    \mathcal{L}_{ssl}=\mathcal{L}_{ssl}^{s}+\mathcal{L}_{ssl}^{e},
\end{gather}
where $M, N$ are the number of students and exercises respectively,
$sim(\cdot)$ is the similarity function which is set to cosine distance,
and $\tau$ is hyperparameter temperature which can mine difficult negative pairs.
\subsection{Training}
Self-supervised graph learning is performed as an auxiliary task, in parallel with the main cognitive diagnosis task.
They are optimized with a multi-task training strategy at the same time.
The total loss of our method is:
\begin{equation}
    \mathcal{L}=\mathcal{L}_{main}+\lambda_{1}\mathcal{L}_{ssl}+\lambda_{2}\|\Theta\|^{2},
\end{equation}
where $\Theta$ represents learnable model parameters.
The Adam optimizer is adopted to minimize $\mathcal{L}$.

\begin{table}[t]
\centering
\caption{Statistics of experimental datasets.}
\label{dataset}
\begin{tabular}{ccc}
\hline
Dataset & junyi & ASSIST \\ \hline
Students & 10000 & 3644 \\
Exercises & 835 & 14439 \\
Knowledge Concepts & 835 & 123 \\
Interactions & 220799 & 281890 \\ 
Interactions per student & 22.07 & 77.36 \\ 
Density & 0.02644 & 0.00536 \\ \hline 
\end{tabular}
\end{table}

\begin{table*}[t]
\centering
\caption{Experimental results on student performance prediction in percentage. The bold indicates the best result.}
\label{result}
\resizebox{\textwidth}{!}{%
\begin{tabular}{cccllccllccllccll}
\multicolumn{17}{c}{(a)ASSIST}                                                                                                                                                                                                                                                                                                                                                 \\ \hline
\multicolumn{1}{c|}{train:test} & \multicolumn{4}{c|}{5:5}                                                               & \multicolumn{4}{c|}{6:4}                                                               & \multicolumn{4}{c|}{7:3}                                                               & \multicolumn{4}{c}{8:2}                                           \\ \hline
\multicolumn{1}{c|}{Methods}    & $ACC$            & $RMSE$           & $ACC_{50}$         & \multicolumn{1}{l|}{$RMSE_{50}$}        & $ACC$            & $RMSE$           & $ACC_{50}$         & \multicolumn{1}{l|}{$RMSE_{50}$}        & $ACC$            & $RMSE$           & $ACC_{50}$         & \multicolumn{1}{l|}{$RMSE_{50}$}        & $ACC$            & $RMSE$           & $ACC_{50}$         & $RMSE_{50}$        \\ \hline
\multicolumn{1}{c|}{IRT}        & 67.90          & 45.82          & 65.74          & \multicolumn{1}{l|}{47.06}          & 68.58          & 45.44          & 66.61          & \multicolumn{1}{l|}{46.99}          & 69.39          & 45.07          & 66.20          & \multicolumn{1}{l|}{47.12}          & 69.73          & 44.69          & 66.83          & 46.73          \\
\multicolumn{1}{c|}{MIRT}       & 70.53          & 48.98          & 69.33          & \multicolumn{1}{l|}{45.25}          & 70.07          & 47.98          & 69.91          & \multicolumn{1}{l|}{45.37}          & 71.38          & 47.08          & 70.21          & \multicolumn{1}{l|}{44.66}          & 72.00          & 46.27          & 70.83          & 45.72          \\
\multicolumn{1}{c|}{NCD}        & 71.88          & 43.73          & 70.18          & \multicolumn{1}{l|}{46.46}          & 72.01          & 43.49          & 71.18          & \multicolumn{1}{l|}{45.11}          & 72.79          & 43.08          & 71.21          & \multicolumn{1}{l|}{44.25}          & 72.77          & 42.74          & 71.77          & 43.82          \\
\multicolumn{1}{c|}{RCD}        & 71.97          & 43.29          & 71.06          & \multicolumn{1}{l|}{41.65}          & 72.14          & 43.12          & 72.06          & \multicolumn{1}{l|}{41.22}          & 72.69          & 42.73          & 72.18          & \multicolumn{1}{l|}{40.81}          & 72.82          & 42.50          & 71.93          & 39.23          \\ \hline
\multicolumn{1}{c|}{SCD}        & \textbf{72.24} & \textbf{43.15} & \textbf{73.00} & \multicolumn{1}{l|}{\textbf{40.94}} & \textbf{72.40} & \textbf{42.93} & \textbf{73.08} & \multicolumn{1}{l|}{\textbf{40.18}} & \textbf{72.88} & \textbf{42.64} & \textbf{73.15} & \multicolumn{1}{l|}{\textbf{39.46}} & \textbf{73.17} & \textbf{42.33} & \textbf{72.94} & \textbf{38.34} \\ \hline
                                &                &                &                &                                     &                &                &                &                                     &                &                &                &                                     &                &                &                &                \\
\multicolumn{17}{c}{(b)junyi}                                                                                                                                                                                                                                                                                                                                                  \\ \hline
\multicolumn{1}{c|}{train:test} & \multicolumn{4}{c|}{5:5}                                                               & \multicolumn{4}{c|}{6:4}                                                               & \multicolumn{4}{c|}{7:3}                                                               & \multicolumn{4}{c}{8:2}                                           \\ \hline
\multicolumn{1}{c|}{Methods}    & $ACC$            & $RMSE$           & $ACC_{50}$         & \multicolumn{1}{l|}{$RMSE_{50}$}        & $ACC$            & $RMSE$           & $ACC_{50}$         & \multicolumn{1}{l|}{$RMSE_{50}$}        & $ACC$            & $RMSE$           & $ACC_{50}$         & \multicolumn{1}{l|}{$RMSE_{50}$}        & $ACC$            & $RMSE$           & $ACC_{50}$         & $RMSE_{50}$        \\ \hline
\multicolumn{1}{c|}{IRT}        & 78.93          & 39.18          & 74.27          & \multicolumn{1}{l|}{44.00}          & 79.13          & 38.87          & 76.31          & \multicolumn{1}{l|}{41.84}          & 79.35          & 38.41          & 77.75          & \multicolumn{1}{l|}{40.37}          & 79.45          & 38.30          & 76.59          & 39.52          \\
\multicolumn{1}{c|}{MIRT}       & 79.91          & 37.72          & 75.86          & \multicolumn{1}{l|}{40.48}          & 80.14          & 37.43          & 76.47          & \multicolumn{1}{l|}{42.20}          & 80.35          & 37.28          & 77.49          & \multicolumn{1}{l|}{40.83}          & 80.54          & 37.22          & 76.93          & 38.34          \\
\multicolumn{1}{c|}{NCD}        & 77.93          & 38.90          & 75.00          & \multicolumn{1}{l|}{40.80}          & 77.42          & 39.97          & 76.02          & \multicolumn{1}{l|}{40.49}          & 78.12          & 38.90          & 77.39          & \multicolumn{1}{l|}{39.61}          & 77.74          & 38.97          & 77.54          & 39.54          \\
\multicolumn{1}{c|}{RCD}        & 80.86          & 36.79          & 77.21          & \multicolumn{1}{l|}{36.65}          & 80.85          & 36.85          & 77.31          & \multicolumn{1}{l|}{35.87}          & 80.95          & 36.63          & 77.83          & \multicolumn{1}{l|}{33.57}          & 81.04          & 36.55          & 77.96          & 32.97          \\ \hline
\multicolumn{1}{c|}{SCD}        & \textbf{81.16} & \textbf{36.69} & \textbf{77.75} & \multicolumn{1}{l|}{\textbf{34.79}} & \textbf{81.00} & \textbf{36.69} & \textbf{77.88} & \multicolumn{1}{l|}{\textbf{34.01}} & \textbf{81.14} & \textbf{36.55} & \textbf{78.41} & \multicolumn{1}{l|}{\textbf{32.95}} & \textbf{81.17} & \textbf{36.54} & \textbf{78.44} & \textbf{31.72} \\ \hline
\end{tabular}%
}
\end{table*}

\section{Experiments}
The experiments are conducted on real-world cognitive diagnostic datasets to answer the following research questions:
\begin{itemize}
    \item
    \textbf{RQ1}: How does SCD perform compared to the state-of-the-art cognitive diagnosis models?
    \item 
     \textbf{RQ2}: How does SCD perform on students with sparse interaction data?
    \item \textbf{RQ3}: What is the advantage of our edge dropout strategy over random strategy?
    \item \textbf{RQ4}: How about the interpretability of SCD diagnostic results?
\end{itemize}
\subsection{Experimental Settings}
\subsubsection{Datasets}
We conduct experiments on two realword datasets: junyi \footnote{\url{https://pslcdatashop.web.cmu.edu/DatasetInfo?datasetId=1198}} and ASSIST \footnote{\url{https://sites.google.com/site/assistmentsdata/home/2009-2010-assistment-data/skill-builder-data-2009-2010}}.
Junyi dataset is collected from the Chinese e-learning website Junyi Academy.
ASSIST is a publicly available dataset collected by ASSISTments online tutoring system for student performance prediction, and we choose the "skill-builder data 2009-2010" version.
Both two datasets contain students' interaction records with exercises and exercises' relation with knowledge concepts.
To demonstrate the ability of SCD on sparse data, we retain students with the number of interaction records above 5.
Detailed statistical information of the processed data is shown in table~\ref{dataset}.
To explore the effect of different sparse data on the experimental results, 
we divided the data set into different proportions.

\subsubsection{Baseline}
We compare our SCD with the following CD models:
\begin{itemize}
    \item \textbf{IRT}~\cite{lord1952theory}. IRT is a basic CD model which leverages a linear function to predict the probability that a student correctly answers a question.
    \item \textbf{MIRT}~\cite{reckase2009multidimensional}. MIRT extends the representation of students and exercises in IRT from one-dimensional to multidimensional.
    \item \textbf{NCD}~\cite{wang2020neural}. NCD introduces nonlinearity and replaces the manually designed prediction function with a neural network.
    \item \textbf{RCD}~\cite{gao2021rcd}. RCD is the state-of-the-art CD model. It introduces relations between knowledge concepts and models the relations using the graph structure.
\end{itemize}

\subsubsection{Hyperparameter Settings and Metrics}
We implement our SCD with PyTorch.
For IRT, MIRT and NCD, we use the code provided by EduCDM~\cite{bigdata2021educdm}.
We re-implement RCD to make it easy to train and achieve the same results as the original code.
For each model we set the batch size to 256.
As for graph-based models, \textit{i.e.} RCD and SCD, we set the layers of the graph network to 2.

We adopt $RMSE$~(Root Mean Square Error) and $ACC$~(Prediction Accuracy) to evaluate the models.
In addition, for the long-tailed problem, we design $RMSE_{50}$ and $ACC_{50}$ to focus on long-tailed students and to avoid the effect of a small number of students with the majority of interaction data on the results.
Specifically, we calculate $RMSE_{50}$ and $ACC_{50}$ by average the $RMSE$ and $ACC$ separately for the top 50\% students with fewer interaction data:

\begin{align}
    RMSE_{50}&=\frac{1}{N_{50}}\sum_{s_{i} \in S_{50}}RMSE(s_{i}) \\
    ACC_{50}&=\frac{1}{N_{50}}\sum_{s_{i} \in S_{50}}ACC(s_{i}), 
\end{align}
where $S_{50}$ are the top 50\% students with fewer interaction data,
and $N_{50}$ is the amount of these students.

\begin{figure}[t]
	\centering
	\begin{subfigure}{\linewidth}
		\centering
		\includegraphics[width=\linewidth]{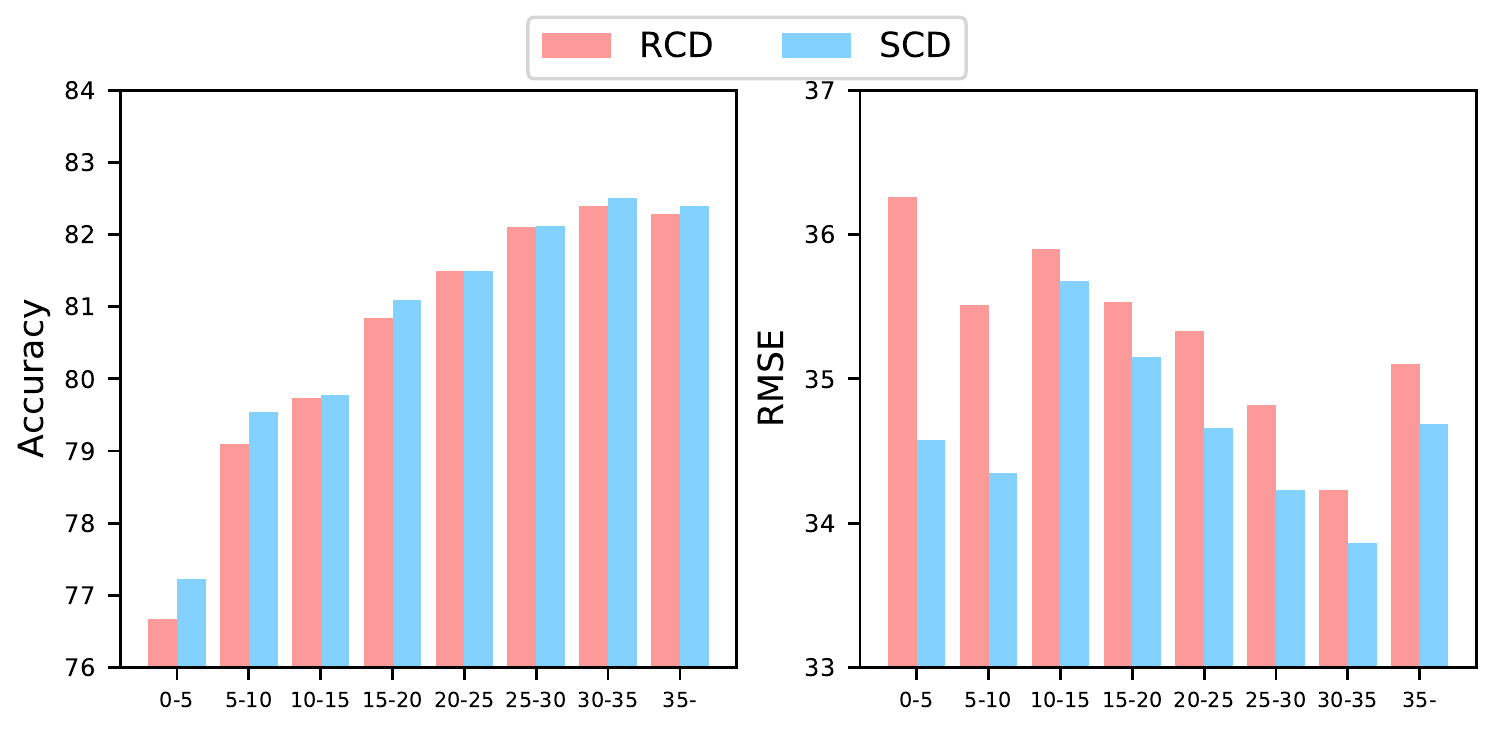}
		\caption{junyi}
		\label{group:a}
	\end{subfigure}
	\centering
	\begin{subfigure}{\linewidth}
		\centering
		\includegraphics[width=\linewidth]{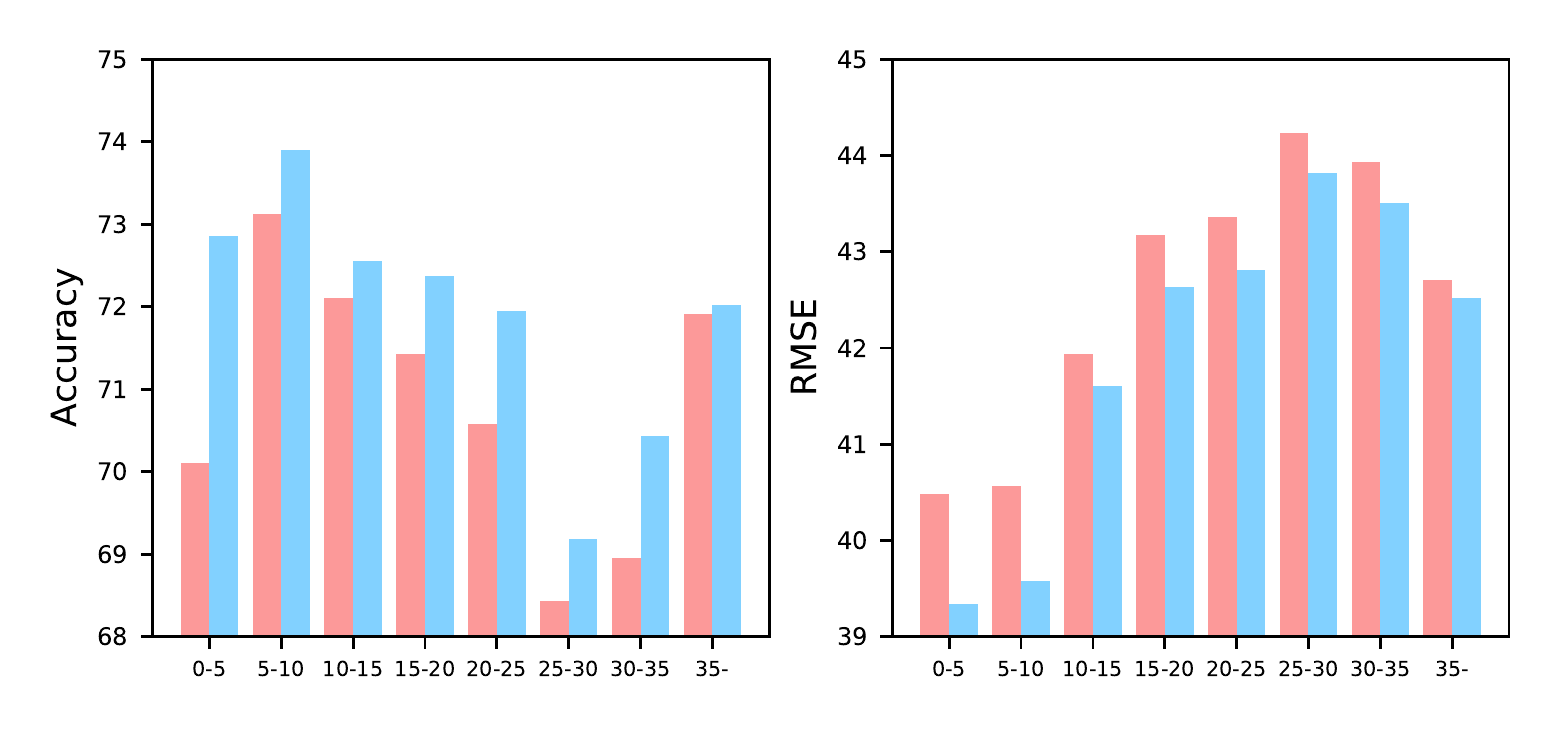}
		\caption{ASSIST}
		\label{group:b}
	\end{subfigure}
	\centering
	\caption{Performance comparison over different student groups. The horizontal coordinate is the number of interactions per student in the group.}
	\label{group}
\end{figure}

\subsection{Performance Comparison (RQ1)}
Table~\ref{result} shows the results comparison between our SCD with four baselines and the best results are shown in bold.
The experimental results show that SCD outperforms all baselines.
Especially, compared with the same graph based method RCD, we find that our SCD outperforms RCD on $ACC$ and $RMSE$ and achieves more significant improvements on $ACC_{50}$ and $RMSE_{50}$.
This indicates that:
\begin{itemize}
    \item There is a serious long-tailed problem in the cognitive diagnostic system, \textit{i.e.}, many students have very little interaction data.
    This long-tailed students would be ignored when evaluating the model by common metrics, where their results will only have a very small impact on the whole results.
    \item Our SCD can focus on the long-tailed students.
    The results on $ACC_{50}$ and $RMSE_{50}$ show that the self-supervised learning auxiliary task can effectively improve the performance of the model on these students.
\end{itemize}

\begin{table}[t]
\centering
\caption{Results of study on edge dropout in percentage. The bold indicates the best results.}
\label{random}
\resizebox{0.5\textwidth}{!}{%
\begin{tabular}{ccccccccc}
\hline
\multicolumn{1}{c|}{Dataset}    & \multicolumn{4}{c|}{ASSIST}                                                            & \multicolumn{4}{c}{junyi}                                         \\ \hline
\multicolumn{1}{c|}{Methods}    & $ACC$            & $RMSE$           & $ACC_{50}$       & \multicolumn{1}{c|}{$RMSE_{50}$}      & $ACC$            & $RMSE$           & $ACC_{50}$       & $RMSE_{50}$      \\ \hline\hline
\multicolumn{9}{c}{5:5}                                                                                                                                                                      \\ \hline
\multicolumn{1}{c|}{SCD-random} & 71.01          & 43.34          & 71.17          & \multicolumn{1}{c|}{41.77}          & 80.82          & 36.91          & 77.12          & 36.83          \\
\multicolumn{1}{c|}{SCD}        & \textbf{72.24} & \textbf{43.15} & \textbf{73.00} & \multicolumn{1}{c|}{\textbf{40.96}} & \textbf{81.16} & \textbf{36.69} & \textbf{77.75} & \textbf{34.79} \\ \hline\hline
\multicolumn{9}{c}{6:4}                                                                                                                                                                      \\ \hline
\multicolumn{1}{c|}{SCD-random} & 72.20          & 43.13          & 72.07          & \multicolumn{1}{c|}{43.20}          & 80.95          & 36.78          & 77.35          & 35.70          \\
\multicolumn{1}{c|}{SCD}        & \textbf{72.40} & \textbf{42.93} & \textbf{73.08} & \multicolumn{1}{c|}{\textbf{40.18}} & \textbf{81.00} & \textbf{36.69} & \textbf{77.88} & \textbf{34.01} \\ \hline\hline
\multicolumn{9}{c}{7:3}                                                                                                                                                                      \\ \hline
\multicolumn{1}{c|}{SCD-random} & 72.63          & 42.71          & 72.04          & \multicolumn{1}{c|}{40.78}          & 81.00          & 36.63          & 77.85          & 33.54          \\
\multicolumn{1}{c|}{SCD}        & \textbf{72.88} & \textbf{42.64} & \textbf{73.15} & \multicolumn{1}{c|}{\textbf{39.46}} & \textbf{81.14} & \textbf{36.55} & \textbf{78.41} & \textbf{32.95} \\ \hline\hline
\multicolumn{9}{c}{8:2}                                                                                                                                                                      \\ \hline
\multicolumn{1}{c|}{SCD-random} & 72.95          & 42.38          & 72.04          & \multicolumn{1}{c|}{39.19}          & 81.10          & 36.57          & 77.97          & 33.02          \\
\multicolumn{1}{c|}{SCD}        & \textbf{73.17} & \textbf{42.33} & \textbf{72.94} & \multicolumn{1}{c|}{\textbf{38.34}} & \textbf{81.17} & \textbf{36.54} & \textbf{78.44} & \textbf{31.72} \\ \hline\hline
\end{tabular}%
}
\end{table}

\subsection{Performance on Long-tailed Students(RQ2)}
To answer RQ2 and explore how SCD benefits from self-supervised learning, we group students according to the number of interactions and measure the performance of RCD and SCD on each group of students separately.
Specifically, we divide the students into nine groups, and the amount of interaction data per student in the group is within a specific range.
The number of students in each group, the total amount of interactions, and the amount of interaction data per student are shown in table~\ref{students}.

Figure~\ref{group} shows the results for grouped students.
We find that SCD achieves a large improvement on groups with a limited number of interactions~(\textit{e.g.}, 0-5, 5-10), while these groups make up the majority of students. 
Furthermore, SCD performs similarly or even better than RCD on groups where the interaction data are not sparse~(\textit{e.g.}, 30-35, 35-).
This indicates that the introduction of self-supervision positively impacts all data.

\subsection{Strategy of Edge Dropout(RQ3)}
In SCD, we design an importance-based edge dropout strategy.
Specifically, we generate the retention probability of each edge based on the degree of its vertices.
To prove the effectiveness of our strategy, we designed SCD-random as a variant:
we retain all edges in the student-exercise interaction subgraph with the same probability $p$, and $p$ is adjusted so that the view generated by SCD-random has a similar number of edges as the view generated from SCD.

Table~\ref{random} shows the comparison between SCD and SCD-random.
SCD outperforms SCD-random in all metrics.
And SCD has a greater advantage on $ACC_{50}$ and $RMSE_{50}$.
From the comparison we can see that SCD-random as a self-supervised learning method does not have an advantage on the long-tailed problem.
This is because completely random strategy may lose essential graph structure information.
An equal-probability dropout of all edges does not help the long-tailed nodes but makes their information more sparse.
Even worse, the nodes can become isolated.
Comparing with SCD-random, the edge drop strategy in our SCD can better enhance the weight of long-tailed nodes in the generated view to improve the performance on long-tailed students.

\begin{table}[t]
\centering
\caption{Statistics of diagnosis case study. The knowledge concepts associated with each exercise are given. \ding{52} indicates that the student correctly answered the exercise, and \ding{56} otherwise. All scores are ground truth, not predicted results.}
\label{case1}
\begin{tabular}{|l|l|l|l|}
\hline
 & Exercise\#1 & Exercise\#2 & Exercise\#3 \\ \hline
Concept & A,B & C & D,E \\ \hline
Student\#1 & \ding{52} & \ding{52} & \ding{52} \\ \hline
Student\#2 & \ding{52} & \ding{56} & \ding{56} \\ \hline
\end{tabular}
\end{table}

\begin{figure}[t]
    \centering
    \includegraphics[width=0.5\textwidth]{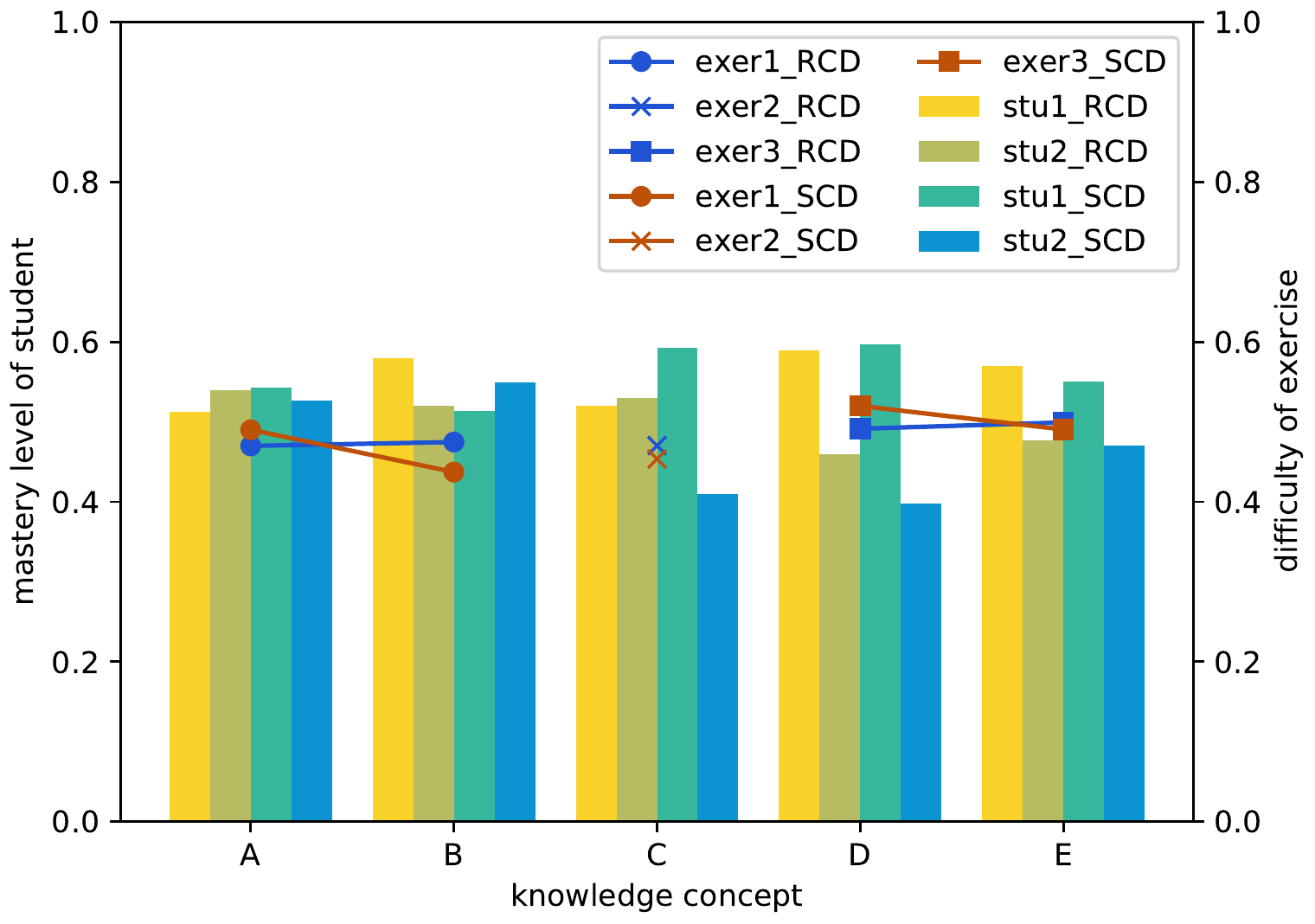}
    \caption{Result of diagnosis case study. The horizontal coordinate is the knowledge concept, the left vertical coordinate is the student's mastery level of the knowledge concept, and the right vertical coordinate is the difficulty of the exercise on the corresponding knowledge concept. The suffixes in the figure legends indicate the diagnosis results of different models.}
    \label{case2}
    \end{figure} 

\subsection{Diagnostic Case Studie(RQ4)}
To answer RQ4, we construct a diagnostic case study.
We chose two students and three exercises that they had answered,
from which student\#2 is from the long-tailed student population.
Table~\ref{case1} shows the statistical information of these students and exercises.
Exercise\#1 is related to Knowledge Concept A\&B,
while exercise\#2 is related to Knowledge Concept C,
and D\&E for exercise\#3.
All students' scores on the exercises have been given.

Figure~\ref{case2} shows the diagnostic results of RCD and SCD.
As for the diagnosis result of SCD,
student\#1
has mastered all the knowledge concepts at a higher level than the difficulty of the corresponding exercises, and all three exercises are correct; 
student\#2 has mastered the knowledge concepts related to exercise\#1, but not exercises\#2 and\#3, so the exercise\#1 is right and it is failed on the others.
RCD has been equally successful in diagnosing Student\#1. 
In the RCD's diagnosis results for Student\#2, he has a higher level of mastery of knowledge concept C than the difficulty of exercise\#2.
This is a failed diagnosis for RCD.
The difference in diagnostic results for long-tailed students demonstrates the strength of SCD's diagnostic capabilities for these students.

\section{Conclusion}
We proposed a self-supervised graph learning framework for cognitive diagnosis~(SCD), which applies contrast loss to provide self-supervised signals for the cognitive diagnosis. It assists the main task to enhance the predictive performance on long-tailed students.
Specifically, we generate different views of the student-exercise graph by discarding the student-exercise relationship edges in a way that retains the important edges.
The consistency of the representation of the same nodes on different sparse views are maximized to make the model focus on long-tailed nodes and provide self-supervised signals.
Experiments on real-world datasets demonstrate the effectiveness on the long-tailed problem and the advantages of our view generation strategy.
We hope to conduct further studies of view generation strategies based on the characteristics of cognitive diagnostic task.

\bibliography{aaai23}

\end{document}